\documentclass[prl,aps,twocolumn,superscriptaddress,balancelastpage]{revtex4-1}
\usepackage{latexsym,amssymb,bm,bbold,amsfonts,amstext,graphicx,bbm,bm,relsize,dsfont,times}

\usepackage[dvipsnames]{xcolor}
\usepackage{amsmath} % AMS Math Package
\usepackage{amsthm} % Theorem Formatting
\usepackage{enumerate} %enumerate
\usepackage{float}

\usepackage[unicode=true, breaklinks=false, pdfborder={0 0 1}, backref=false, colorlinks=true, linkcolor=blue, citecolor=blue]{hyperref}

\usepackage[linesnumbered,ruled,vlined]{algorithm2e}
\usepackage{algcompatible,amsmath}
\algnewcommand\INPUT{\item[\textbf{Input:}]}%
\algnewcommand\OUTPUT{\item[\textbf{Output:}]}%

\newcommand{\tr}{\mbox{tr}}
\def\bra#1{\langle{#1}|}
\def\ket#1{|{#1}\rangle}

\newcommand{\ketbra}[2]{\ket{#1}\!\bra{#2}}

\def\BraVert{\egroup\,\mid\,\bgroup}

\usepackage{tcolorbox}
\definecolor{yellow}{HTML}{FFDB25}
\definecolor{white}{HTML}{FFFFFF}
\definecolor{light_yellow}{HTML}{FFF6B3}
\definecolor{green}{HTML}{00BA82}
\definecolor{mag}{HTML}{D33EBE}
\definecolor{light_orange}{HTML}{FFC7A5}
\definecolor{red}{HTML}{FF006C}
\definecolor{blue}{HTML}{0046FF}
\definecolor{light_green}{HTML}{A5FFD2}
\definecolor{light_blue}{HTML}{A5ECFF}
\definecolor{light_deep_blue}{HTML}{D4E2FF}
\definecolor{az}{HTML}{00B9E6}

\begin{document}

\title{An algorithm for solving unconstrained unitary quantum brachistochrone problems}

\author{Francesco Campaioli}
\email{francesco.campaioli@monash.edu}
\affiliation{School of Physics and Astronomy, Monash University, Victoria 3800, Australia}

\author{William Sloan}
\affiliation{School of Physics and Astronomy, Monash University, Victoria 3800, Australia}

\author{Kavan Modi}
\affiliation{School of Physics and Astronomy, Monash University, Victoria 3800, Australia}

\author{Felix A. Pollock}
\affiliation{School of Physics and Astronomy, Monash University, Victoria 3800, Australia}

\date{\today}

\begin{abstract}
We introduce an iterative method to search for time-optimal Hamiltonians that drive a quantum system between two arbitrary, and in general mixed, quantum states. The method is based on the idea of progressively improving the efficiency of an initial, randomly chosen, Hamiltonian, by reducing its components that do not actively contribute to driving the system. We show that our method converges rapidly even for large dimensional systems, and that its solutions saturate any attainable bound for the minimal time of evolution. We provide a rigorous geometric interpretation of the iterative method by exploiting an isomorphism between geometric phases acquired by the system along a path and the Hamiltonian that generates it, and discuss resulting similarities with Grover's quantum search algorithm.
Our method is directly applicable as a powerful tool for state preparation and gate design problems.
\end{abstract}

\maketitle

\makeatletter

\textbf{Introduction} --- Inspired by the problem of finding the curve of fastest descent between two points, quantum brachistochrone problems (QBPs) aim to find the Hamiltonian that generates the time-optimal evolution between two given quantum states. Such problems have been considered, in various forms, to obtain accurate  minimum-time protocols for the control of quantum systems~\cite{Wang2011, Geng2016}, clarify the role of entanglement and quantum correlations in time-optimal evolution~\cite{Borras2008, Basilewitsch2017}, study the speed of Hermitian and non-Hermitian Hamiltonians~\cite{Bender2007, Assis2008, Mostafazadeh2007}, and improve bounds on the minimal time of evolution, known as quantum speed limits~\cite{Russell2017, Brody2015, Campaioli2018}. QBPs have a special role in quantum information theory and technology, where they are of fundamental importance to accurately perform tasks such as preparing a desired state of a system, or implementing a specific gate, while satisfying the strict physical and fault tolerance requirements imposed by the locality of interactions and short decoherence times~\cite{Caneva2009, Wang2015}. For this reason, their solutions have found applications in information processing~\cite{Werschnik2007,Reich2012, Jager2014,Dur2014,Huang2014,Schulte-Herbruggen2005}, quantum state preparation~\cite{Vandersypen2005, VanFrank2016, Islam2011, Senko2015, Jurcevic2014a, Zhou2016}, cooling~\cite{Wang2011,Machnes2012}, metrology~\cite{Kessler2014,Giovannetti2004c,Burkard1999}, and quantum thermodynamics~\cite{Rezakhani2009,DelCampo2013b,Binder2015,Campaioli2017}. Moreover, QBPs give a physical interpretation to the complexity of quantum algorithms, which emerges from the minimisation of the time required to obtain the desired unitary transformation~\cite{Carlini2006}.

Solving QBPs is generally hard, and accurate analytic and numerical solutions are only known for some special cases, such as the unconstrained unitary evolution of pure states, and control problems of well structured quantum systems~\cite{Carlini2006, Levitin, Carlini2011, Carlini2012, Carlini2013, Brody2015, Wang2015}. There are methods that can be used to address a relatively large class of QBPs, but obtaining precise solutions becomes increasingly challenging as the dimension of the system grows, and the constraints on the  control Hamiltonian become more complex. For instance, the quantum brachistochrone equations proposed in Ref.~\cite{Carlini2006} involve the solution of ordinary differential equations with boundary conditions, for which there are no efficient numerical methods when high-dimensional systems are considered~\cite{krotov1993, Carlini2007, Rezakhani2009, Carlini2008, Wang2015}. One crucial open problem is the case of unconstrained unitary evolution between two mixed states. As opposed to the case of pure states, the solution to this QBP is not known, except for special cases with highly degenerate spectra. This is akin to the problem of finding tight bounds on the minimal time of evolution, known as quantum speed limits (QSLs). These well-known bounds are attainable for pure states, but are often loose for mixed states due to the complicated structure of the space of density operators~\cite{Deffner2013, DelCampo2013, Sun2015}, though we have recently proposed geometric methods to derive simple, efficient, and tight bounds for the time of evolution of mixed states~\cite{Campaioli2018, Campaioli2018b}.

In this Letter we take a similar approach to solving the complementary problem, that of finding the optimal unconstrained unitary evolution between two mixed states. That is, we look for the generator $H$ of a unitary operator $U = \exp(-i H t)$ that takes a mixed state $\rho$ to $\sigma = U \ \rho \ U^\dag$, such that the transition time $t$ is minimised. We introduce an iterative method to search for the optimal time-independent Hamiltonian, while respecting an energetic constraint. The method progressively improves the efficiency of the Hamiltonian~\cite{Uzdin2012} that generates the evolution, by applying an algorithm that rapidly converges even for large dimensional systems. 

We investigate the efficacy of our method using bounds on the minimal time of evolution, i.e., QSLs. By comparing the achievable upper bound, provided by the iterative method, with the inviolable lower bound offered by QSLs, we demonstrate the strong synergy of the two results: When the two are found to be close, we can be sure that both results are  close to optimal. We study the performance of our method with respect to the size of the system, and its dependence on parameters such as the convergence threshold, before discussing direct applications of the algorithm, its potential combination with time-optimal gate design and Monte-Carlo methods, and its geometric interpretation, juxtaposing it to that of Grover's quantum search algorithm.

%%%%%%%%%%%%%%%%%%%%%%%%
\textbf{Time-optimal evolution and Hamiltonian efficiency} --- Let us begin by considering the QBP for a two-level system, defined by the initial state $\rho = (\mathbb{1}+ p \, \Lambda_x)/2$, and the target state $\sigma = (\mathbb{1}+ p \, \Lambda_y)/2$. Here $p \in (0,1]$, and $\bm{\Lambda}=(\Lambda_x,\Lambda_y,\Lambda_z)$ are Pauli matrices. The Bloch vectors for the two states are given by $\bm{r} = (p,0,0)$ and $\bm{s} = (0,p,0)$, respectively. Any unitary $O(\bm{\varphi}) = e^{i\varphi_1}\ketbra{s_1}{r_1}+e^{i\varphi_2} \ketbra{s_2}{r_2}$ can be used to map $\rho\to\sigma$, where $\ket{r_k}$ and $\ket{s_k}$ are the eigenvectors of $\rho$ and $\sigma$, respectively, and where $\bm{\varphi} = (\varphi_1,\varphi_2)$ represent the different geometric phases that the state gathers from $O(\bm{\varphi})$. The Hamiltonians $H(\bm{\varphi}) = i \log O(\bm{\varphi})$ defines different evolutions depending on the choice of $\bm{\varphi}$. 

Let us focus on two possible choices for these phases, $\bm{\varphi}_z = (\pi/4,\pi/4)$, and $\bm{\varphi}_{xy} = (\pi/4,-3\pi/4)$. The associated Hamiltonians are given by $H(\bm{\varphi}_z) = \Lambda_z \pi/4$, and $H(\bm{\varphi}_{xy}) = (\Lambda_x+\Lambda_y) \sqrt{2}\pi/4$. These two Hamiltonians generate different paths on the Bloch sphere. Under energetic constraints, such as fixing the standard deviation $\Delta H_\rho = \sqrt{\tr{\rho H^2} - \tr{\rho H}^2}$ with respect to the initial state, the lengths of these paths meaningfully correspond to evolution times. The path generated by $H(\bm{\varphi}_z)$ connects $\rho$ to $\sigma$ with an arc of great circle, i.e., the geodesic with respect to the Fubini-Study (FS) metric~\footnote{The FS metric is the unique, unitarily invariant metric on the space of pure states~\cite{Bengtsson2008}}, thus it constitutes a solution to the considered QBP for any homogeneous energy constraint~\cite{Russell2017}, whereas $H(\bm{\varphi}_{xy})$ draws a \emph{longer} path, which is not time optimal.

A heuristic explanation for the variable performance of different Hamiltonians is that the Bloch vectors $\bm{h}$ associated with them $H = \bm{h}\cdot \bm{\Lambda}$ have a different orientation with respect to $\bm{r}$ and $\bm{s}$. In particular, when $\bm{h}$ is orthogonal to $\bm{r}$, it generates a rotation on a plane that passes through the origin of the Bloch sphere, while when $\bm{h}$ is not perpendicular to $\bm{r}$, it generates a rotation on a plane that does not. The slower evolution can be seen as due to a less efficient use of the resources encoded in constraints on the energy, which are \emph{wasted} on parts of the Hamiltonian that not actively drive the system~\cite{Uzdin2012}. The notion of efficiency can be formalised for pure states as in Ref.~\cite{Uzdin2015}, and generalised to mixed states, as discussed in the Supplemental Material.

This intuitive geometric argument can be generalised for dimension $d>2$ by replacing the notion of orthogonality between vectors with commutation relations between states and Hamiltonians. Given an operator $\rho$, the space of Hamiltonians (the Lie algebra $\mathfrak{su}(d)$) splits into a `parallel' subspace, closed under multiplication, that commutes with $\rho$, and an orthogonal `perpendicular' subspace that does not. This allows us to decompose the Hamiltonian $H$ into components $H_\|$ and $H_\perp$ which are elements, respectively, of these two subspaces, such that $[H_\|,\rho]=0$ and $[H,\rho]=[H_\perp,\rho]$~\footnote{These correspond to elements of the vertical and horizontal tangent bundles in a fibre bundle representation where the state space (with a given spectrum) is the base manifold and the phases $\mathbf{\varphi}$ are the fibres.}. This observation leads to the idea at the core of the iterative method that we will introduce in the next section, where efficient Hamiltonians for a QBP are achieved by requiring their parallel component to be vanishing at all times during the evolution.

\textbf{Iterative method for efficient Hamiltonians}\label{p:method} --- We can now consider the more general QBP defined by an arbitrary isospectral pair of initial and final states $\rho =\sum_k \lambda_k\ketbra{r_k}{r_k}$ and $\sigma =\sum_k \lambda_k\ketbra{s_k}{s_k}$ of finite dimension $d$. When non-degenerate states are considered, the operator
\begin{equation}
    \label{eq:o_phi}
    O(\bm{\varphi}) = \sum_k e^{i\varphi_k}\ketbra{s_k}{r_k},
\end{equation}
represents the most general unitary that connects initial and final states $\rho$ and $\sigma$, while $\varphi_k$ represent the geometric phases gained evolving along the path generated by $i\log O(\bm{\varphi})$. When degenerate states are considered, the phases $\varphi_k$  are replaced by unitary operators $U_{k} \in SU(m)$ on the subspace associated with eigenvalues $\lambda_{k}$ with multiplicity $m$. Since any unitary $O(\bm{\varphi})$ maps $\rho\to\sigma$, we can choose an arbitrary initial phase $\bm{\varphi}^{(0)}$ to obtain the initial unitary $O^{(0)} = O(\bm{\varphi}^{(0)})$. The Hamiltonian $H^{(0)} = i \log O^{(0)}$, canonically associated with $O(\bm{\varphi}^{(0)})$, is then split into a parallel and perpendicular component $\mathcal{M}_\rho[H^{(0)}] = H^{(0)}_\|$ and $H^{(0)}_\perp = H^{(0)}-H^{(0)}_\|$, such that $[H^{(0)}_\|,\rho]=0$, via a map $\mathcal{M}_\rho$ that depends on $\rho$, and which will be referred to as \emph{mask}. A possible choice for such a mask is given by
\begin{align}
\label{eq:non_degenerate_mask}
    &\mathcal{M}_\rho[H]=D \big( M \circ D^\dagger H D) \big) D^\dagger, 
\;\mbox{with} \; M_{ij} = \delta_{\lambda_i\lambda_j},
\end{align}
where $D$ diagonalises $\rho$, i.e., its columns are given by the eigenstates $\{\ket{r_k}\}$ of $\rho$, and where $\circ$ is the Hadamard, i.e., element-wise, product \cite{Horn2012}. Note that, when degenerate states are considered, the eigenvalues' multiplicity defines the structure of the mask via $\delta_{\lambda_i\lambda_j}$.
\begin{algorithm}[t]
\caption{Iterative method for efficient Hamiltonians}
\label{a:algorithm}
\begin{algorithmic}[1]
    \INPUT Initial state $\rho = \sum_{k=1}^d \lambda_k \ketbra{r_k}{r_k}$ and final state $\sigma = \sum_{k=1}^d \lambda_k \ketbra{s_k}{s_k}$. Initial phase $\bm{\varphi}^{(0)}$.
    Threshold for convergence $\varepsilon$.
    \OUTPUT Optimal Hamiltonian.
    \STATE Initialise $O^{(0)}=\sum _k \ketbra{s_k}{r_k}$, $H^{(0)}=i \log O^{(0)}$, $H_\parallel ^{(0)}=\mathcal{M}_\rho[H^{(0)}]$; 
    \WHILE{$\|H_\parallel ^{(j)} \|_\mathrm{HS}>\varepsilon \| H^{(j)}\|_\mathrm{HS}$} \\
    Set $O^{(j)}=O^{(j-1)}e^{i H_\parallel ^{(j-1)}}$, $H^{(j)}=i \log O^{(j)}$, $H_\parallel ^{(j)}=\mathcal{M}_\rho[H^{(j)}]$;
    \ENDWHILE $\;\;$ 
    \STATE \textbf{return} the final Hamiltonian $H^{(n)}$ of the sequence $\{H^{(j)}\}_{j=0}^n$.
\end{algorithmic}
\end{algorithm}

We now notice that the unitary 
$U^{(0)} = \exp[-i H_\|^{(0)}]$ can be composed with $O^{(0)}$ to obtain another unitary $O^{(1)} = O^{(0)} U^{(0)}$, formally equivalent to evolving with the time-dependent Hamiltonian $e^{-iH^{(0)}t}H^{(0)}_\perp e^{iH^{(0)}t}$ (as we detail in the Supplemental Material), which also drives $\rho\to\sigma$ since $[\rho,H_\|]=0$. In general, the unitary $O^{(1)}$ is associated with a new geometric phase, and the Hamiltonian $H^{(1)} = i \log O^{(1)}$ draws a different path, which might be shorter (or longer) than that generated by $H^{(0)}$. In a second iteration, we apply the mask to the new Hamiltonian to obtain $H^{(1)}_\|=\mathcal{M}_\rho[H^{(1)}]$, and define another gate $O^{(2)}$ analogously. By iterating this method we obtain a sequence of Hamiltonians $\{H^{(j)}\}_{j=0}^n$ that drive $\rho\to\sigma$ via $\exp[-i H^{(j)}]$. The unitary at each step is related to the previous one by
\begin{align}
    \label{eq:gate_n}
    & O^{(j+1)} =O^{(j)} e^{-i \mathcal{M}_\rho[H^{(j)}]}, \; \mbox{with} \; H^{(j)} = i \log O^{(j)}.
\end{align}
A necessary and sufficient condition for the routine to reach a \emph{fixed point} $H^{(n)}$, such that $H^{(n+1)}=H^{(n)}$, is given by $\mathcal{M}_\rho[H^{(n)}]=0$, i.e., when the parallel part of the $n$-th Hamiltonian vanishes under the action of the mask $\mathcal{M}_\rho$, which follows trivially from the fact that in this case $e^{-i\mathcal{M}_\rho[H^{(n)}]}=\mathbb{1}$.

When implementing the method numerically one has to fix a convergence threshold to stop the routine as soon as the parallel component $H_\|^{(n)}$ becomes small enough with respect to the full Hamiltonian $H^{(n)}$.
We have chosen to quantify this threshold with a positive number $\varepsilon\ll 1$, such that the method stops when~$\lVert H_\|^{(n)}\rVert_{HS}\leq\varepsilon \lVert H^{(n)}\rVert_{HS}$. Accordingly, the number of iterations $n$ required for the method to converge implicitly depends on $\varepsilon$. Numerical evidence suggests that the sequence always converges towards a Hamiltonian $H^{(n)}$ that is fully \textit{perpendicular} with respect to $\rho$ along the whole evolution, in the sense that the parallel components eventually vanish within the precision defined by $\varepsilon$. This iterative method is summarised in its simplest form by Algorithm~\ref{a:algorithm}, and can be interpreted as an optimisation of energy cost associated with the different geometric phases $\bm{\varphi}$, accomplished via the recursive suppression of ineffective components of the Hamiltonians.

There is also a geometric interpretation of our method analogous to that of Grover's famous quantum search algorithm~\cite{Grover1997}. The latter aims to find the unique input, encoded in a quantum state, of a function that produces a particular output value. Its action on an initialisation state $\ket{\psi^{(0)}}$ can be interpreted as a sequence of rotations that converges with high probability to the desired state. Similarly, the iterative method introduced here can be seen as a sequence of rotations of some initialisation unit vector $\hat{\bm{h}}^{(0)}$, associated with the Hamiltonian via $H^{(0)}=\bm{h}^{(0)}\cdot \bm{\Lambda}$, where $\bm{\Lambda} = (\Lambda_1,\dots,\Lambda_{d^2-1})$ forms a Lie Algebra for $SU(d)$. However, unlike Grover's algorithm, the vectors $\{\bm{h}^{(n)}\}$ do not span a two-dimensional real plane, but a high-dimensional subspace of $\mathbb{R}^{d^2-1}$. 

\begin{figure}[t]
    \centering
    \includegraphics[width=0.49\textwidth]{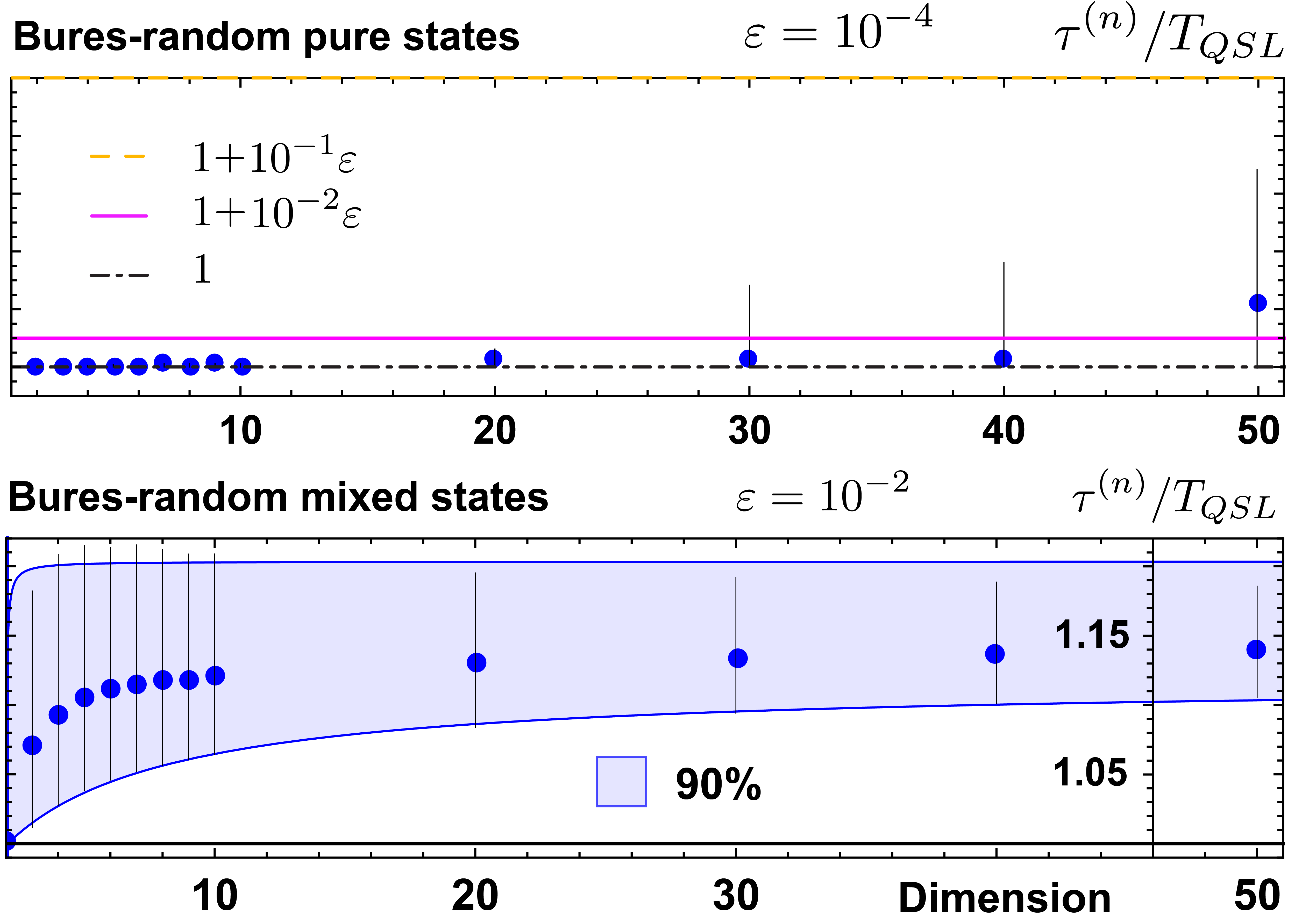}
    \caption{\textbf{Performance of the method} --- The performance of the method is here studied for Bures-random pure (top) and mixed (bottom) states, using the ratio $\tau^{(n)}/T_{QSL}\geq 1$ between the optimised time $\tau^{(n)}$ required to drive $\rho\to\sigma$ with the solution $H^{(n)}$, and the quantum speed limit $T_{QSL}$ given in Ref.~\cite{Campaioli2018}, evaluated for each pair of states $\rho$, $\sigma$, and optimised Hamiltonian $H^{(n)}$. The plotted points show the average performance for different Hilbert space dimensions $d$, with the \emph{error bars} representing $99\%$ (top) and $90\%$ (bottom) confidence intervals. The shaded area (bottom) represents a fit of the $90\%$ confidence interval for the average $\tau^{(n)}$. The convergence threshold is $\varepsilon=10^{-4}$ for pure states (top), and $\varepsilon=10^{-2}$ for mixed states (bottom). The sample size is $10^4$ for each dimension. For  pure states, the method returns solutions that converge on the QSL as the precision $\varepsilon^{-1}$ is increased (at the expense of requiring more iterations).}
    \label{fig:perf}
\end{figure}

\textbf{Performance of the method} --- 
As anticipated in the introduction, our iterative method can be directly applied to solve QBPs defined by the unconstrained time-optimal unitary evolution of any isospectral pair of density operators. A sub-class of these problems with known solutions are those of unconstrained unitary evolution between pure states or between mixed states whose eigenvalues are all degenerate except one. To demonstrate the performance of our method, we have tested it on Bures-random pairs of pure states $\ketbra{\psi}{\psi}$, $\ketbra{\phi}{\phi}$ of dimension $d=2,\dots,100$, successfully obtaining Hamiltonians that are time-optimal and fully efficient with respect to the notion of efficiency introduced in Ref.~\cite{Uzdin2015}. We can be confident that the solutions we obtain are globally optimal by exploiting the fact that the quantum speed limit for unitary evolution of pure states is attainable~\cite{Levitin}. Comparing the evolution time $\tau^{(n)}$ of the optimized Hamiltonian $H^{(n)}$ obtained from Algorithm~\ref{a:algorithm} with the minimal evolution time given by the Mandelstam-Tamm bound $T_{QSL}(\ket{\psi},\ket{\phi}) = d_{FS}(\ket{\psi},\ket{\phi})/\Delta H^{(n)}$~\cite{Deffner2017}, where $d_{FS}(\ket{\psi},\ket{\phi})=\cos^{-1}(|\bra{\psi}\ket{\phi}|)$ is the FS~distance, we find $\tau^{(n)} \approx T_{QSL}$, within the precision imposed by $\varepsilon$, for all cases, as shown in Fig.~\ref{fig:perf}.

A more challenging test was run on random pairs of mixed states of dimension $d=3,\dots,100$ (for initial states with both degenerate and non-degenerate spectra), for which a general solution to the unconstrained unitary QBP is not known. Since the quantum speed limit for the unitary evolution of mixed states~\cite{Campaioli2018} is in general not tight, it is harder to benchmark the quality of the solutions provided by our method in this case. On the other hand, the natural synergy between this iterative method and the QSLs introduced in Ref.~\cite{Campaioli2018} can be used to gauge the performance of the  former and the tightness of the latter. The \emph{actual} minimal time of evolution $\tau_{\text{min}}$ for a given choice of $\rho$ and $\sigma$ is bounded as
\begin{gather}
    \label{eq:bound_on_t_actual}
    T_{QSL} \leq \tau_{\text{min}} \leq \tau^{(n)},
\end{gather}
where $T_{QSL}$ corresponds to the inviolable lower bound offered by QSLs of Ref.~\cite{Campaioli2018}, and $\tau^{(n)}$ to an achievable upper bound, provided by the solutions obtained with our iterative method. 
The first inequality can be saturated for pure states and for states with $d-1$ degenerate eigenvalues. For such case, the second inequality can be saturated up to the precision imposed by $\varepsilon$, due to the fact that the optimized time $\tau^{(n)}$ inherently depends on the convergence threshold $\varepsilon$, which can be seen as an implicit trade-off between $n$ and $\varepsilon$. By looking at the difference between $\tau^{(n)}$ and $T_{QSL}$ for the given solution $H^{(n)}$, we can assess the quality of both results, which we find to coincide within the precision set by $\varepsilon$ for some choices of $\rho$ and $\sigma$, even when their degeneracy structure differs from that of pure states.

Moreover, the performance of the solutions $H^{(n)}$ obtained with this iterative method is stable under small perturbations of the initial state $\rho$. These could be introduced through convex mixing $\rho\to\rho' = (1-\delta) \rho + \delta \: \chi$, where $\chi$ is a random state of the same dimension as $\rho$, or unitary rotation $\rho\to\rho' = e^{i V \delta} \rho \; e^{-i V \delta}$, where $V$ is a random Hamiltonian of unit Hilbert-Schmidt norm. As shown in the Supplemental Material, the difference between the solution to the original problem and that for the perturbed version does not grow quickly as a function of  the perturbation strength $\delta$. We now discuss the rate at which the algorithm converges.

\begin{figure}[t]
    \centering
    \includegraphics[width=0.49\textwidth]{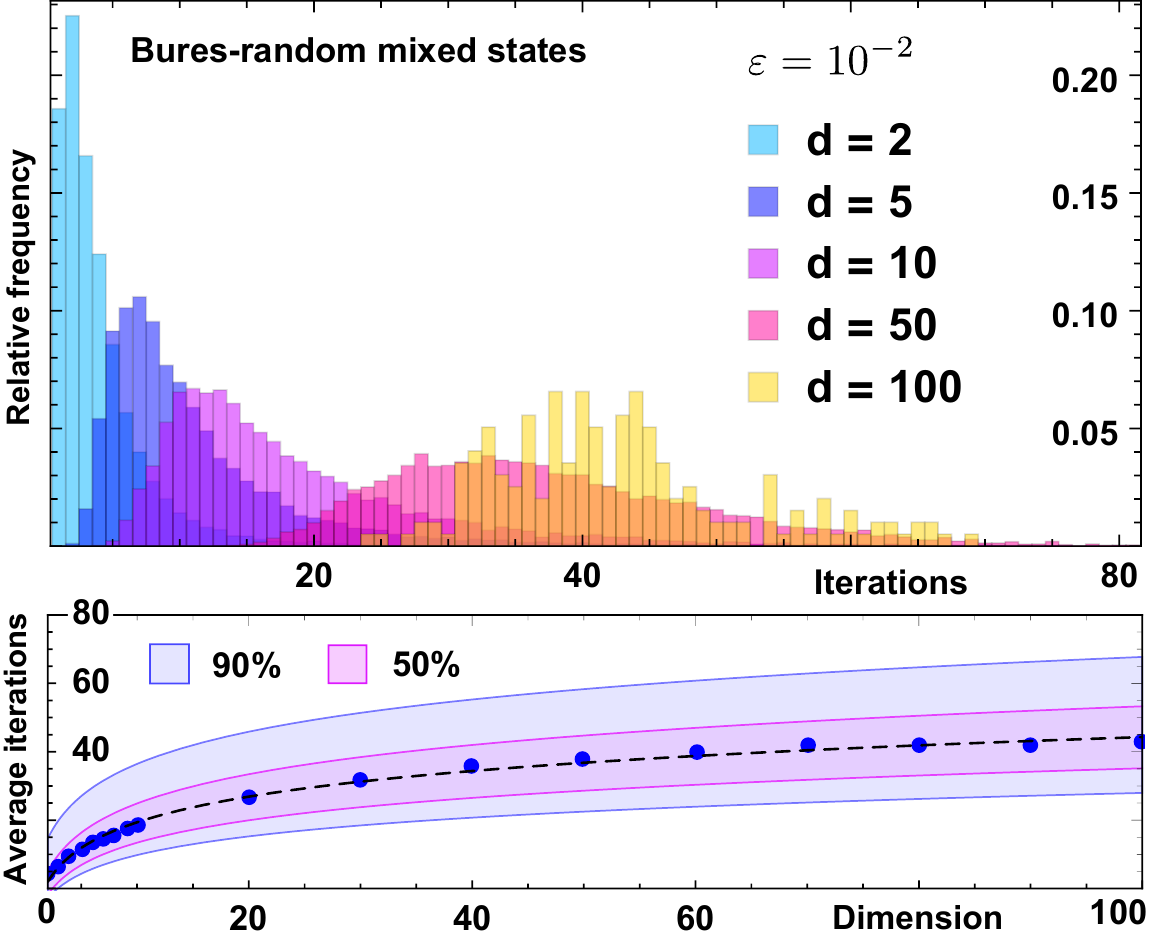}
    \caption{\textbf{Number of iterations required for convergence} --- The number $n$ of iterations required for convergence is represented above for Bures-random mixed initial and final states $\rho$ and $\sigma$, with Hilbert space dimension $d=2,\dots,100$, as indicated in the plots. The histograms (top) show the relative frequency of the number of iterations required for convergence. The plotted points (bottom) show the average number of iterations required for convergence for the considered system dimensions, together with their $50\%$ and $90\%$ confidence regions. The black dashed line represents a logarithmic fit of the average number of iterations required to converge as a function of $d$. The sample size is $10^4$ for each dimension $d\leq50$, and $200$ for $d>50$, while the convergence threshold is chosen to be $\varepsilon=10^{-2}$.}
    \label{fig:iter}
\end{figure}

\textbf{Convergence of the method} --- 
The number $n$ of iterations required for convergence generally grows with the dimension $d$ of the system, and the strictness of the precision set by $\varepsilon$. Recalling that the number of elements of the matrices associated with non-commuting density operators $\rho$ and $\sigma$ grows quadratically with $d$, one might expect conservatively that the average number of iterations $\bar{n}$ would grow in the same way. However, $\bar{n}$ grows logarithmically with $d$, as shown in Fig.~\ref{fig:iter}, with a slower growth for the case of pure states and highly degenerate mixed states. Equivalently, for composite systems $\bar{n}$ grows linearly with the number of constituent subsystems.

The choice of initial phase vector $\bm{\varphi}^{(0)}$ also affects the number of iterations required to converge, without noticeably affecting the performance of the end point $H^{(n)}$; we have never observed the case where two different choices for $\bm{\varphi}^{(0)}$ lead to converged solutions with inconsistent evolution times. 
Moreover, while $\bar{n}$ grows with $d$ and $\varepsilon^{-1}$, a good choice of initial geometric phase $\bm{\varphi}^{(0)}$ can still lead to quick convergence, returning runs that can take less then 10 iterations even for $d=100$ and $\varepsilon = 10^{-4}$.
For these reasons, there is the potential to combine Algorithm~\ref{a:algorithm} with Monte-Carlo sampling methods~\cite{Hastings1970,Vanderbilt1984,Kalos2008}, in order to deploy many parallel runs of the iterative method, for the same pair of initial and final states $\rho$ and $\sigma$, with each run initialised with a different, randomly chosen geometric phase $\bm{\varphi}$. Since the whole computation is stopped as soon as one of these runs converges to a solution, the number $n$ of iterations is guaranteed to be smaller than the average one. We give evidence in the Supplemental Material that this approach is feasible. There we show that, for fixed $\rho$ and $\sigma$, the fraction of initial phases that converges quickly is still almost always significant, even with large $d$ and $\varepsilon^{-1}$. The use of Monte-Carlo methods comes at the expense of requiring a much larger number of simultaneous tasks, which form an \textit{embarrassingly parallel workload}~\cite{Herlihy2012}.

\textbf{Discussion} --- 
In this Letter, we have introduced an iterative algorithm to obtain efficient time-independent Hamiltonians that generate fast evolution between two $d$-dimensional isospectral states $\rho$ and $\sigma$. This constitutes a solution to the quantum brachistochrone problems for unitary evolution of mixed states, with direct applications to state preparation and quantum optimal control problems. 
We have shown that a necessary and sufficient condition for our method to converge is to reach an efficient Hamiltonian, and provided extensive numerical evidence that our method rapidly reaches a solution even for large Hilbert space dimension $d\sim 100$. In particular, we have shown that the average number of iterations required for convergence grows logarithmically with the dimension of the system and that this number can be further reduced by running multiple instances of the algorithm in parallel. For this reason our method provides a viable and powerful approach to the design of time-optimal Hamiltonians of high-dimensional systems. 

A possible application of the iterative method introduced here could be found in its combination with quantum optimal control methods. Algorithm~\ref{a:algorithm} can be used to identify a unitary $\widetilde{O}$, associated with an efficient Hamiltonian $\widetilde{H}$, to be implemented by means of time-optimal gate design methods, such as those experimentally realised in Ref.~\cite{Geng2016}. As opposed to QBPs defined by the specific choice of a pair of state $\rho$ and $\sigma$, time-optimal gate design problems aim to realize a certain unitary $U$ (independently of the input states) by means of a time-dependent Hamiltonian, achieved by optimizing some control pulses. More generally, our method has broad application in further elucidating the geometric structure of quantum control, since it has been shown that quantum brachistochrone problems can be recast as those of finding geodesics in the space of unitary operators~\cite{Wang2015}.

Our method could also be extended to more general control problems. For instance, a similar approach could be taken to the open evolution quantum systems, where the Hamiltonian is replaced by a Liouvillian generating a family of CPTP maps. Moreover, whether open or not, constraints on the form of the generators could be included. Such constraints are naturally imposed by physical restrictions on the order and range of interactions, and can dramatically change the bounds on minimal evolution time, as recently shown in Ref.~\cite{Campaioli2017}. Even for unconstrained unitary evolution, there may be alternative formulations of our algorithm which perform better under certain circumstances. In the Supplemental Material we briefly outline a version (which does not qualitatively differ in its performance) where the part of the Hamiltonian that commutes with the \emph{final state} $\sigma$ is additionally removed in each iteration. It is possible that there are further modifications along these lines that could improve the rate of convergence on the solution of the QBP. Finally, the study of the stability of these methods on experimental errors, such as the precision on the control parameters and on time measurements, would be of paramount practical importance.

% Bibliography ------------------------------
\bibliography{main.bbl}

\clearpage
\newpage
\begin{center}
    \textbf{\large Supplemental Material}
\end{center}

\section{Derivation of Algorithm~\ref{a:algorithm}}
\label{sm:derivation}
With reference to the problem defined in the second section of this Letter, let $O=\sum_k\ketbra{s_k}{r_k}$, and $i \log O=H=H_\|+H_\perp$, where $H_\|$ commutes with $\rho$. Consider the unitary $U_t$ generated by the time dependent Hamiltonian $O_t H_\perp O^\dagger_t$, where $O_t = \exp[-i H t]$, which satisfies the equation
\begin{equation}
    \label{eq:partial_derivative_U}
    \partial_t U_t = -i O_t H_\perp O^\dagger_t U_t.
\end{equation}
To leading order in time, this will rotate the perpendicular part of the Hamiltonian to follow the system's evolution. Now consider the unitary $\widetilde{U}_t=O^\dagger_t U_t$, whose equation of motion is given by
\begin{equation}
    \label{eq:eq_of_motion_U}
    \begin{split}
    \partial_t \widetilde{U}_t &= -i H_\perp \widetilde{U}_t
 + (\partial_t O^\dagger_t)U_t \\
 & = -i(H_\perp - H)\widetilde{U}_t \\
 & = i H_\| \widetilde{U}_t.
 \end{split}
\end{equation}
Eq.~\eqref{eq:eq_of_motion_U} implies that $\widetilde{U}_t = \exp[i H_\|t]$, thus that $\widetilde{U}_t$ commutes with $\rho$. Accordingly, $U=U_1=O_1\widetilde{U}_1$ transforms $\rho\to\sigma$, just as well as $O$ does. The procedure can be iterated until it reaches a fixed point, i.e., until $H = H_\perp$ and $H_\| = 0$.

The iterative method introduced here involves removing the part of the Hamiltonian operator that commutes with the initial state $\rho$. A similar approach could be taken, where the part of the Hamiltonian that commutes with the final state $\sigma$ is removed, and the unitary joining the two states is acted on from the left in each iteration, instead of from the right. It also possible to remove both components simultaneously, replacing Eq.~\eqref{eq:gate_n} with $O^{(j+1)}=e^{i\mathcal{M}_{\sigma}[H^{(j)}]}O^{(j)}e^{-i\mathcal{M}_{\rho}[H^{(j)}]}$. While we find this approach generally changes the convergence rate of the algorithm, it does not seem to perform significantly better than the algorithm outlined in the main text.

\section{Hamiltonian Efficiency for Density Operators}
\label{sm:efficiency}      

When pure states are considered, we refer to the the notion of Hamiltonian efficiency introduced in Ref.~\cite{Uzdin2015}, given by $\eta[H,\rho] := \Delta H_\rho / \lVert H\rVert_{op}$, with $\lVert \cdot\rVert_{op}$ being the operator norm. 
It is possible to consider different energetic constraints, such as $\lVert H \rVert_{HS} = \omega$, known as finite energy bandwidth, or $\lVert H\rVert_{op} =\omega$, which corresponds to bounding the largest eigenvalue of the Hamiltonians. Different notions of efficiency can be obtained for any choice of speed and total energy measure. The efficiency measures how much of the energy available, quantified by $\lVert H\rVert_{op}$, is transferred to the system to 
drive the evolution, and thus \textit{converted} into speed $\Delta H_\rho$. 
If we assess the performance of the two Hamiltonians $H(\bm{\varphi}_z)$ and $H(\bm{\varphi}_{xy})$ considered in the Letter using $\eta$, we obtain $\eta(H(\bm{\varphi}_z),\rho)=p$, and $\eta(H(\bm{\varphi}_{xy}),\rho)=p/2$. The dependence on $p$ (and thus on the purity of the states) implies that this notion of efficiency cannot not be saturated in general. 

\begin{figure}[t]
    \centering
    \includegraphics[width=0.48\textwidth]{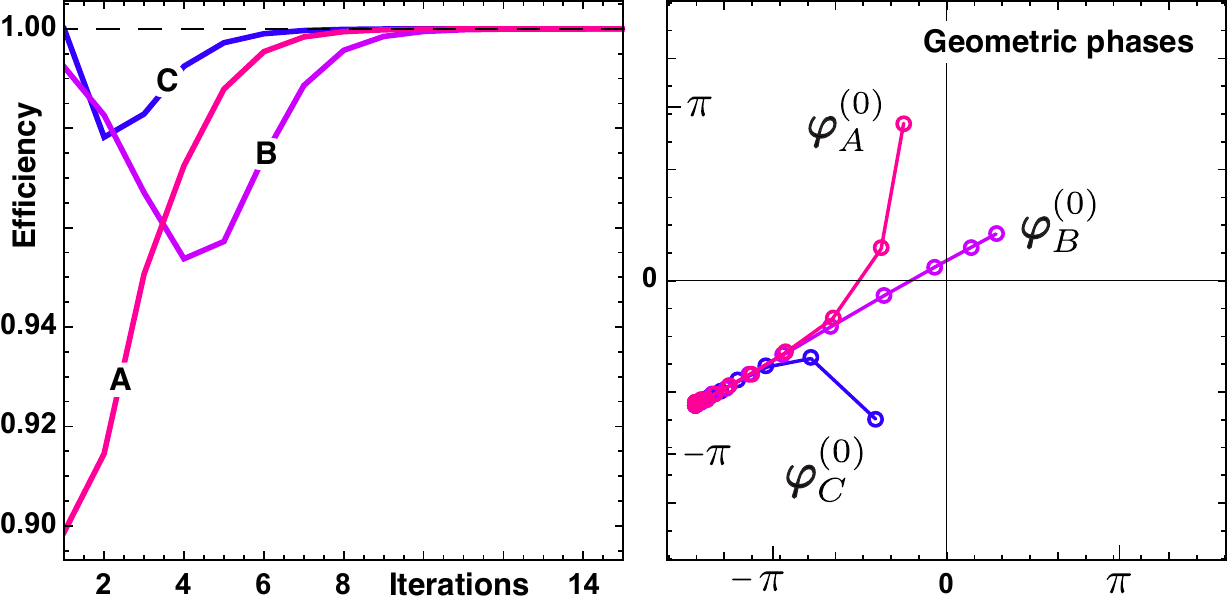}
    \caption{\textbf{Non-monotonic solutions and their geometric phase trajectories} --- Algorithm~\ref{a:algorithm} is here applied on the same choice of $\rho$ and $\sigma$ for three different choices of initial geometric phases $\bm{\varphi}^{(0)}_A$, $\bm{\varphi}^{(0)}_B$, and $\bm{\varphi}^{(0)}_C$. The Hamiltonian efficiency $\eta^\star(H^{(j)},\rho)$ of the Hamiltonians in the three different sequences is plotted along the iterations $i$ for each run (left). Since each Hamiltonian (and unitary) has an associated geometric phase as prescribed by Eq.~\eqref{eq:o_phi}, we can plot the trajectory of the geometric phase for each run, such as as $\bm{\varphi}^{(j)}_A=(\varphi^{(j)}_{A,1},\varphi^{(j)}_{A,2},\varphi^{(j)}_{A,3})$ for run $A$. By neglecting the first phase, which can be absorbed into a global phase, we plot the remaining components on a plane (right). All runs converge to the same solutions, however, while the efficiency of run $A$ converges monotonically, the others converge non-monotonically. The trajectories of their geometric phases do not give away the signature of non-monotonicity, and straight path can correspond to the slowest non-monotonic descent towards the fixed point. Moreover, such trajectories can show richness and variety for different choices of $\rho$ and $\sigma$, and of initial geometric phases.} 
    \label{fig:non_mono}
\end{figure}

A possible notion of Hamiltonian efficiency that can be saturated for the case of density operators is given by
\begin{gather}
    \label{eq:efficiency_star}
    \eta^\star(H,\rho) = \frac{\sqrt{\tr[\rho^2H^2]-\tr[(\rho H)^2]}}{\sqrt{\tr[\rho^2 H^2]-(\tr[(\rho H)^2]-\tr[\rho H]^2)}},
\end{gather}
which is a positive function smaller than 1, that is saturated for $H_\| = 0$. The numerator of $\eta^\star$ is proportional to the speed $\lVert \dot{\rho}\rVert_{HS}$ of the generator $\dot{\rho} = -i[H,\rho]$~\cite{DelCampo2013,Deffner2013,Campaioli2018}, and reduces to $\Delta H_\rho$ when pure states are considered, for which $\eta^\star =1 \leftrightarrow \eta =1$. Moreover, like $\Delta H_\rho$, $\lVert \dot{\rho}\rVert_{HS}$ is also invariant under the addition of a parallel component $H_\|$ with respect to $\rho$. Accordingly, the Hamiltonians generated by the iterative method are of unit efficiency for all the considered QBPs, which reflect the ability to converge to a fully-perpendicular Hamiltonian. However, the convergence of this iterative method is, in general, non-monotonic with respect to this notion of efficiency, which can decrease before reaching $\eta^\star =1$ after several iterations, as shown in Fig.~\ref{fig:non_mono}.

\section{Performance and convergence of Algorithm~\ref{a:algorithm}}
\label{sm:performance}
The two parameters that we considered to study the performance of the iterative method introduced in this Letter are $n$ and $\tau^{(n)}/T_{QSL}$, respectively given by the number of iterations required to converge, and the ratio between the optimised time $\tau^{(n)}$ and the QSL for the problem defined by the chosen initial and final state and $H^{(n)}$. Eq.~\ref{eq:bound_on_t_actual} implies that $\tau^{(n)}/T_{QSL} \geq 1$, but since the $T_{QSL}$ is not tight in general, $\tau^{(n)}/T_{QSL} > 1$ does not imply that $H^{(n)}$ is not a time-optimal Hamiltonian.

To study how the performance of the iterative method varies with $d$, we sampled a large number of pairs $\rho,\sigma$ from the Bures-ensemble of pure and mixed states of dimension up to 100. We have considered a sample size of $10^4$ pairs of states $\rho$ and $\sigma$ for dimension $2,3,4,5,6,7,8,9,10,20,30,40,50$, and a sample of $10^2$ pairs for $d=60,70,80,90,100$.
Numerical evidence suggests that the average number of iterations required for convergence $\bar{n}$ grows linearly with $\varepsilon^{-1}$, and, as shown in Fig.~\ref{fig:iter}, logarithmically with $d$. 

\begin{figure}[t]
    \centering
    \includegraphics[width=0.48\textwidth]{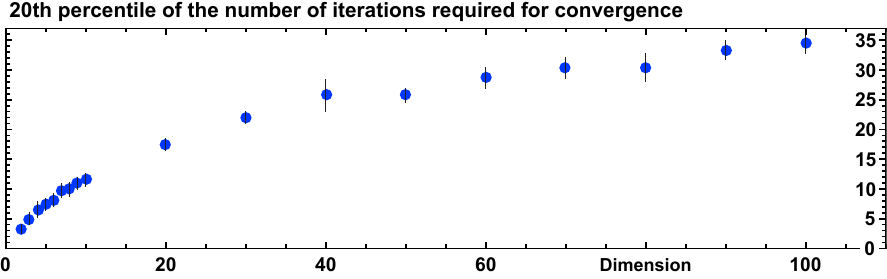}
    \caption{\textbf{Growth of the 20th percentile of $n$} --- The average 20th percentile of the number of iterations required for convergence is plotted along the Hilbert space dimension $d$, with the \emph{error bars} representing the standard deviation. For each dimension we have sampled $10^2$ pairs of initial and final states $\rho$ and $\sigma$, while sampling $10^2$ random initial phases for each pair. The slow growth of the lower end of the distribution of $n$ suggests that Monte-Carlo sampling methods are a feasible option to speed up the computation even for large $d$.}
    \label{fig:monte_carlo_worthy}
\end{figure}

The choice of initial geometric phase $\bm{\varphi}^{(0)}$ strongly affects $n$ without affecting $\tau^{(n)}$ (within the precision set by $\varepsilon$). We studied the dependence of $n$ on the choice of initial geometric phases by running several instances of the method for the same pair of states $\rho$ and $\sigma$, uniformly sampling each phase $\varphi^{(0)}_j$ in the interval $[0,2\pi]$. As shown in Fig.~\ref{fig:monte_carlo_worthy}, the slow growth of the 20th percentile of the number of iterations required for converges suggests that Monte-Carlo sampling methods could combined with this method to speed up the computation.

\section{Stability under perturbations}
\label{sm:stability}
\begin{figure}[t]
    \centering
    \includegraphics[width=0.48\textwidth]{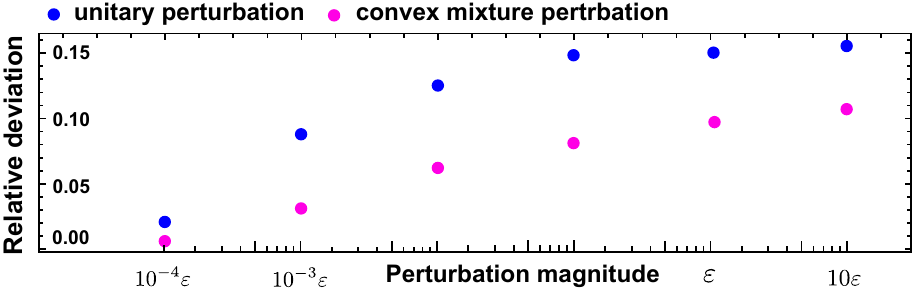}
    \caption{\textbf{Stability under perturbations} --- The relative deviation from the unperturbed solutions, given by $\lVert H^{(n)} - {H'}^{(m)} \ \rVert_{HS}/\lVert H^{(n)} \rVert_{HS}$,where ${H'}^{(m)}$ is the solutions to the perturbed problem, is plotted against the perturbation strength $\delta$, with $\varepsilon=10^{-3}$. Numerical evidence suggest that the relative deviations grow slowly as a function of the perturbation strength, and are negligible for $\delta \ll \varepsilon$.}
    \label{fig:perturbation}
\end{figure}

As mentioned in the Letter, the iterative method defined by Algorithm~\ref{a:algorithm} is stable under small perturbations of the initial and final states $\rho$ and $\sigma$. We considered convex and unitary perturbations, respectively given by 
\begin{align}
    \label{eq:convex_mixture}
    &\rho\to\rho' = (1-\delta) \rho + \delta \; \chi, \text{and} \\
    \label{eq:unitary_transform}
    &\rho\to\rho' = e^{i V \delta} \rho \; e^{-i V \delta},
\end{align}
where $\chi$ is a random state of the same dimension as $\rho$, and $V$ is a random Hamiltonian of unit Hilbert-Schmidt norm. For each pair $\rho,\sigma$, the perturbed final state $\sigma'$ is then obtained by applying any unitary $O$ that maps $\{\ket{r_k}\}_{k=1}^d\to\{\ket{s_k}\}_{k=1}^d$ on the perturbed initial state $\rho'$. We look for the effect of perturbations calculating the relative deviations from the unperturbed solutions, given by $\lVert H^{(n)} - {H'}^{(m)} \ \rVert_{HS}/\lVert H^{(n)} \rVert_{HS}$, where ${H'}^{(m)}$ is the solutions to the perturbed problem. Numerical evidence suggest that the relative deviations grow slowly as a function of the perturbation strength $\delta$, and are negligible for $\delta \ll \varepsilon$, as shown in Fig.~\ref{fig:perturbation}.

\end{document}